# Magnetic field dependence of vortex activation energy: a comparison between $MgB_2$, $NbSe_2$ and $Bi_2Sr_2Ca_2Cu_3O_{10}$ superconductors

S. D. Kaushik[1], V. Braccini[2] and S. Patnaik[1]

1 School of Physical Sciences, Jawaharlal Nehru University, New Delhi 110067, India.

2 CNR-INFM LAMIA, C.so, Perrone 24, 16152, Genova, Italy

Corresponding author e-mail: spatnaik@mail.jnu.ac.in

# Abstract

The dissipative mechanism at low current density is compared in three different classes of superconductors. This is achieved by measurement of resistance as a function of temperature and magnetic field in clean polycrystalline samples of $NbSe_2$, $MgB_2$ and $Bi_2Sr_2Ca_2Cu_3O_{10}$ superconductors. Thermally activated flux flow behavior is clearly identified in bulk $MgB_2$. While the activation energy at low fields for $MgB_2$ is comparable to $Bi_2Sr_2Ca_2Cu_3O_{10}$, its field dependence follows a parabolic behavior unlike a power law dependence seen in $Bi_2Sr_2Ca_2Cu_3O_{10}$. We analyze our results based on the Kramer's scaling for grain boundary pinning in $MgB_2$ and $NbSe_2$.

**PACS No.:** 74. 25.Qt, 74.70.Ad, 74.78.Db

**Keywords:** Vortex state, Arrhenius plot, Vortex activation energy, Resistive broadening.

**Introduction**

Technological applicability of superconductors depends on several parameters such as high transition temperature, high upper critical field, low anisotropy and the ability to sustain high current densities at elevated temperatures. In spite of higher transition temperature ($T_c$) in copper oxide based superconductors (with upper critical field in the excess of 150 Tesla), practically the entire superconducting magnet industry centers on liquid helium cooled niobium based superconductors. This is because the in-field critical current density, that is the maximum current density before the onset of dissipation at a given field, is severely suppressed in high $T_c$ materials. Particularly disconcerting is the appearance of resistive behavior at extremely small current densities at all finite temperatures [1]. This happens because of thermally activated hopping of vortex bundles (TAFF) across the pinning sites even if the Lorentz force due to the transport current is much smaller compared to the pinning force at the defect sites. In this respect the advent of two band superconductor magnesium diboride with $T_c \sim 39$ K [2], anisotropy $\gamma \sim 3$ [3] and upper critical field $H_{c2}(0) \sim 60$ T [4] holds promise, especially in conjunction with liquid cryogen free close cycle refrigerators [5]. Similar considerations require that bulk $MgB_2$ must be robust against lossy transport at $\sim 20$ K, the base temperature of single stage cryocoolers.

The clear signature of the thermally activated vortex dynamics is the broadening of superconducting transition in the presence of external magnetic field. However this is not a sufficient condition because the broadening can also be attributed to several other reasons. It may simply be caused by inhomogeneous microstructure with different phases having different transition temperatures. This aspect can be easily verified by the transition width at zero external fields. There are some reports that suggest that the broadening can be due to surface and multiband superconductivity [6]. But in general terms, it is understood that the vulnerability to thermal fluctuations in superconductors is well characterized by the Ginzburg number (Gi) =

$(k_B T_c/H_c(0)\xi_c\xi^2_{ab})^2/2$, where $\xi_{ab}$, $\xi_c$ are coherence lengths for H perpendicular to ab plane and c-axis respectively. The estimated value of Gi for BSCCO is of the order of $10^{-2}$ whereas that for NbSe$_2$ is reported as low as $10^{-8}$ [7]. For MgB$_2$, Angst et al. have separately suggested a value of $10^{-5}$ which is in the intermediate range [8]. We choose NbSe$_2$ for comparison because it is a low T$_c$ superconductor and has electronic anisotropy that is comparable to MgB$_2$. Recently it has also been discussed as a multigap superconductor [9]. In essence our approach is to study and compare the lossy transport in three classes of superconductors in a material form that is closest to long length cables. First we seek to answer whether thermally activated flux flow theories are applicable to polycrystalline MgB$_2$. While TAFF has been reported in thin films of MgB$_2$ [10], in this paper we clearly identify this in polycrystalline MgB$_2$. In spite of large variation in Gi, we observe TAFF behavior in all three superconductors. Our results show that the TAFF activation energy $U_0$ (T,B) follows a parabolic dependence with magnetic field for MgB$_2$ and NbSe$_2$ where as it exhibits power law decay in the case of the cuprate BSCCO. The parabolic dependence is interpreted as a consequence of grain boundary pinning. Further, the magnitude of activation energy at low fields is comparable in BSCCO and NbSe$_2$ but it is an order of magnitude higher in MgB$_2$.

**Experiment**

Polycrystalline samples of MgB$_2$ and NbSe$_2$, and highly c-axis oriented platelets of BSCCO, extracted form silver sheathed tapes, are used in the study. The preparation methods of these samples have been reported elsewhere [11, 12]. The dimension of these samples are 1×3×0.5 mm$^3$, 2×5×0.12 mm$^3$ and 2×4×0.2 mm$^3$ for MgB$_2$ bulk, BSCCO and NbSe$_2$, respectively. Silver sheath of BSCCO tape was removed by etching with NH$_4$OH and H$_2$O$_2$ solution. Linear four-probe technique was used for resistance and I-V measurement. The external field was varied in the range from 0 - 7 T and the direction of transport probe current was kept perpendicular to the applied field. Care was taken in selecting the correct magnitude of probe current so that it followed Ohmic behavior in normal state in each case and yet did not lead to pronounced sample heating. For bulk $MgB_2$ and $NbSe_2$ current was chosen to be 20 mA, and for BSCCO it was 15 mA. All measurements were performed in warming cycle after zero field cooling. The data were taken in a "Cryogenic" 8 Tesla cryogen free magnet system with an attached variable temperature insert.

**Results and discussion**

Figure 1 shows the normalized resistivity as a function of reduced temperature at various applied magnetic field for $MgB_2$ bulk, BSCCO, and $NbSe_2$. The inset in each layer shows the zero field resistivity as a function of temperature and marks the superconducting transition temperature. We also note that the temperature dependence of normal state resistivity in the three samples is entirely different. For $NbSe_2$ the features of charge density wave at T ~ 35 K is clearly identified [13]. On the other hand $MgB_2$ shows robust metallic behavior from room temperature to 40 K. The $T_c$ is found to be 110 K for BSCCO, 39 K for intermetallic polycrystalline $MgB_2$, and 7.5 K for $NbSe_2$. The zero field transition width for $MgB_2$ is ~0.45 K, for $NbSe_2$ it is ~ 0.15 K while for BSCCO it is ~ 1.5 K. The residual resistance ratio ($\rho_{(T=300K)} / \rho_{(T=Tc)}$) is estimated to be 4.3, 14.4, and 12.7 for BSCCO, $MgB_2$ bulk, and $NbSe_2$ respectively. High RRR values point towards relative clean limit of the sample. The numbers are lower than single crystal values and are attributed to the presence of grain boundary defects in polycrystalline samples. Recently Rowell et al. have defined $\rho_{300 K} - \rho_{Tc}$ instead of RRR as the true parameter for determination of intergrain connectivity [14]. The values so calculated, normalized to the respective room temperature resistivity, are found to be 0.76, 0.97, and 0.92 for BSCCO, $MgB_2$ bulk, and $NbSe_2$ respectively. Therefore the bulk samples have similar intergrain connectivity and thus are suitable for comparison of vortex activation energy. We also note that all the samples in the present study show significant broadening in transition width that increases with application of magnetic field. This is most pronounced in the case of BSCCO. In high $T_c$ BSCCO, the magneto-resistance curves near $T_c$ merge together where as both in $MgB_2$ and $NbSe_2$ bulk we observe noticeable shift of the onset with the application of field.

While there have been some reports indicating flux creep from magnetization studies in polycrystalline $MgB_2$ [15], it is still an open question if it has TAFF origin. By definition, TAFF is observed in a narrow window of small current density above the irreversibility line where the log ρ versus log E curve shows a linear dependence in contrast to a weak non linear behavior at higher current densities which is explained by Bardeen – Stephan Flux flow model [16]. This curvature change is the essential evidence for the TAFF model to be applicable. Shown in figure 2 is the log ρ versus log E curve for the $MgB_2$ bulk sample at $\mu_o B = 3$ T. As the temperature increases, there is a clear transition from a highly nonlinear behavior below $T < T_{irr}$ (~28.25 K) to a regime of weakly nonlinear behavior with finite resistance above $T > T_{irr}$. The weakly nonlinear behavior has two regimes. At low currents, the resistivity is due to TAFF where as at higher current it is due to flux flow resistivity given by $\rho_F = \rho_n B / B_{c2}$, where $\rho_n$ is the normal state resistivity. In the TAFF regime, the resistivity is given by [1]

$$\rho_{TAFF}(T, B) = (\nu_0 \phi_o^2 L_c / k_B T) \exp(-U_o(T,B)/k_B T) = \rho_o \exp(-U_o(T,B)/k_B T) \qquad (1)$$

Here $\nu_0$ is the attempt frequency for the hopping of flux bundles (~$10^{12}$ Hz), $\phi_o$ is the flux quantum ($2 \times 10^{-7}$ Gcm$^2$), $L_c$ is the length of the vortex bundle and $U_o$ is the vortex activation energy. In the amorphous limit, the activation energy is given as $U_0 = (J_c \times B) V_c r_p = (J_c \times B) a_0^2 L_c \xi$, where $V_c$ is volume of the flux bundle $R^2_c L_c$, with $R_c$ ~ intervotex distance $a_0$, and $r_p \sim \xi$ that is the range of pinning potential [1]. In figure 3 we show the Arrhenius plots of resistivity in the logarithmic scale versus inverse of temperature. A linear regime is seen in all the four samples with the most pronounced effects in BSCCO. The magnitude of activation energy $U_o$ is directly related to the slope of a straight line drawn on Arrhenius curves in the region of resistivity lower than 10 % of normal state resistivity. From Fig.3, the extrapolated value of $\rho_o$ for $MgB_2$ is $2.38 \times 10^3$ $\mu\Omega$-cm, for BSCCO it is $2.5 \times 10^3$ $\mu\Omega$-cm and for $NbSe_2$ this value is $2.9 \times 10^5$ $\mu\Omega$-cm. These values are more than two orders of magnitude higher than the normal state resistivity value $\rho_n$ (at $T_c$).

The dependence of activation energy on magnetic field has attracted considerable attention in the recent past [1]. In Figure 4 we plot the activation energy $U_o$, determined from the magnetoresistance study versus applied magnetic field. Following Thompson et al.[17], we fit the data using the equation

$$U(B) = aB^\gamma (1-B/B^*)^\delta \qquad (2)$$

Where B is the applied magnetic field, $B^*$ is the irreversibility field, and a, $\gamma$, $\delta$ are scaling parameters. Here the component $(1-B/B^*)^\delta$ represents suppression of superconducting property due to the application of magnetic field and the component $B^\gamma$ represents the variation due to decreasing critical current density. For t < 0.9 Equation 2 fits well for $MgB_2$ sample with the fitting parameter $\gamma = -0.39$, $\delta = 2.1$ with irreversibility field $B^* = 9.67$ T and a = 3410 K. The goodness of the fit $R^2 = 0.95$. On fitting U versus H curve in $MgB_2$ we find a clear signature of parabolic behavior. The fitting parameters and the value of $U_o$ is of the same order as has been calculated for $MgB_2$ polycrystalline thin films by Thompson and our previous report [17, 10]. Thus the activation energy in bulk sample is of the same order as compared to polycrystalline films. Shown in figure 4b is a similar analysis for BSCCO. The best fit follows power law dependence and U goes as $H^{-0.57}$ in the small external magnetic field up to 2 T and changes to $H^{-0.76}$ at higher fields. This crossover to a slow decay in activation energy at high fields has been reported in BSCCO 2223 single crystals [18]. Further, the magnitude of $U_0$ at low fields is an order of magnitude higher in our BSCCO tape platelets as compared to single crystals. The high value of n ($U \sim H^{-\eta}$) and $U_0$ in powder in tube processed BSCCO tape is assigned to improved pinning due to granularity, and defects. We note that like BSCCO, in $MgB_2$ the onset of TAFF behavior occurs at temperature far below $T_c$. For $NbSe_2$ with equation 2, and $\delta = 2$ the fitting parameter are found to be a = 631.70 K, $\gamma$ = -1.1, irreversibility field B* = 14.45 T and the parameter of goodness of the fitting $R^2 = 0.98$. It suggests that in $NbSe_2$ U versus H follows the behavior like that for polycrystalline $MgB_2$. The low field activation energy for $NbSe_2$ is of the same order as BSCCO. We also note that in the low field region below 2 T, the activation energy is $NbSe_2$ decreases faster as compared to $MgB_2$. In intermetallic superconductors, the intergrain pinning is the most dominant flux pinning mechanism because the average grain boundary is far wider than inter vortex distance. The flux pinning force in this regime is given by Kramer's scaling law that is $F_P = J_c \times B = (B/B^*)(1 - B/B^*)^2$ [19,20]. Therefore we conclude that the parabolic field dependence of activation energy as observed in polycrystalline $MgB_2$, and $NbSe_2$ is attributable to strong grain boundary pinning.

The TAFF regime, in the presence of anisotropy, thermal fluctuation, and granularity, has been the focus of many studies [7]. The three parameters that define the TAFF behavior are the i) thermal energy kT, ii) correlated vortex-bundle volume (related to anisotropy) and iii) effectiveness of grain boundary pinning (related to coherence length). We note that while the anisotropy of all the three superconductors are comparable [3, 21], there is a vast difference between the coherence length and thermal energy kT. While the large coherence length and the size of the as grown point defects in $MgB_2$ enables strong grain boundary pinning, the larger anisotropy leads to greater $U_0$ at low fields. The magnitude of $U_0$ at low fields is given by $U_0 \sim \phi_0^2 \xi / 48 \pi^2 \lambda^2$ [17]. For $MgB_2$ using the value $\xi \sim 5$ nm and $\lambda \sim 100$ nm [22] $U_o$ is estimated to be $\sim 10000$ K. Similarly for BSCCO using the value $\xi \sim 0.09$ nm and $\lambda \sim 194$ nm [23] $U_o$ is found to be ~ 7400 K where as for the case of NbSe$_2$ with $\xi$ ~ 2.3 nm and $\lambda$ ~ 230 nm, U$_o$ is calculated to be ~ 1000K. This explains our data very well. Thus we find that while the vulnerability to thermal fluctuation is characterized by 1/$\xi_{ab}\xi_c$, the magnitude of U$_0$ at low field goes as $\xi/\lambda^2$.

To gain further insight into the pinning properties of bulk MgB$_2$, in figure 5 we replot the $\rho$ – J graph of figure 2 with scaling parameters as applicable to vortex glass theory [24]. Our objective is to see whether the vortices in MgB$_2$ behave as 3 D lines or do they decouple into 2D pancakes as seen in BSCCO superconductors [25]. The VGT predicts a change in curvature for log V versus log I curves at a temperature defined as the vortex glass transition T$_g$. The exponents for such universal scaling mechanism give information about the dimensionality of vortices on superconducting systems. The scaled I – V curves at H = 3 T show the curvature change at T$_g$ = 28.8 K. The fitting parameters are found to be z = 4.6 and $\nu$ = 1.05. These values correspond to 3D vortices in our MgB$_2$ samples.

In conclusion, we identify thermally assisted flux flow regime in clean polycrystalline MgB$_2$. The calculated activation energy for NbSe$_2$ and MgB$_2$ follows a parabolic field dependence that is analyzed in the general framework of grain boundary pinning. C-axis oriented BSCCO platelets on the other hand follow a power law dependence H$^{-\eta}$ with $\eta$ ~ 0.57 in the low magnetic field up to ~ 1 T. We also find that at low fields the activation energy U$_0$ relates to the ratio $\xi/\lambda^2$ where as its vulnerability to thermal fluctuations is decided by the ratio 1/$\xi_{ab}^2\xi_c$.

**Acknowledgement**

We thank J. Jiang and I. Naik for the BSCCO and NbSe$_2$ samples and very useful discussions. SBP thanks Department of Science of Technology, India and SDK thanks Council of Scientific and Industrial Research, India for financial support.

**Figure Captions:**

**Figure 1:** Normalized resistivity is plotted against reduced temperature at constant magnetic field for sample under study. Upper layer shows data for $MgB_2$ at the magnetic field 0, 0.1, 0.25, 0.5, 0.75, 1, 1.5, 2, 3, 4, 5, 6 T. Normalized magneto-resistance for BSCCO at magnetic field 0, 0.1, 0.25, 0.5, 1, 2, 3, 4, 5 T is shown in middle layer. Magneto-resistance for $NbSe_2$ at magnetic field 0, 0.1, 0.25, 0.5, 0.75, 1, 1.5, 2, 3, 4, 5 and 6 T is shown in lower layer. Inset in each layer depicts zero field resistivity from room temperature to low temperature.

**Figure 2:** Resistivity is plotted as a function of current density under the influence of external magnetic field of 3 T at constant temperatures.

**Figure 3:** Resistivity on the log scale is plotted against the inverse of temperature at varying magnetic field for $MgB_2$, BSCCO and $NbSe_2$ in respective layers. Arrhenius behavior is observed (see text).

**Figure 4:** Vortex activation energy $U_0$ is plotted against the field H for $MgB_2$, BSCCO and $NbSe_2$ on log – log scale. Joining lines are curve fitted with the relation mentioned in the figures.

**Figure 5:** Current density and resistivity behavior has been scaled to test vortex glass behavior. Scaled current density $J_{sc} = J/T(1-T/T_g)^{2\upsilon}$ is plotted as a function of scaled resistivity $\rho_{sc} = \rho/(1-T/T_g)^{\upsilon(z-1)}$.

**Figure 1**

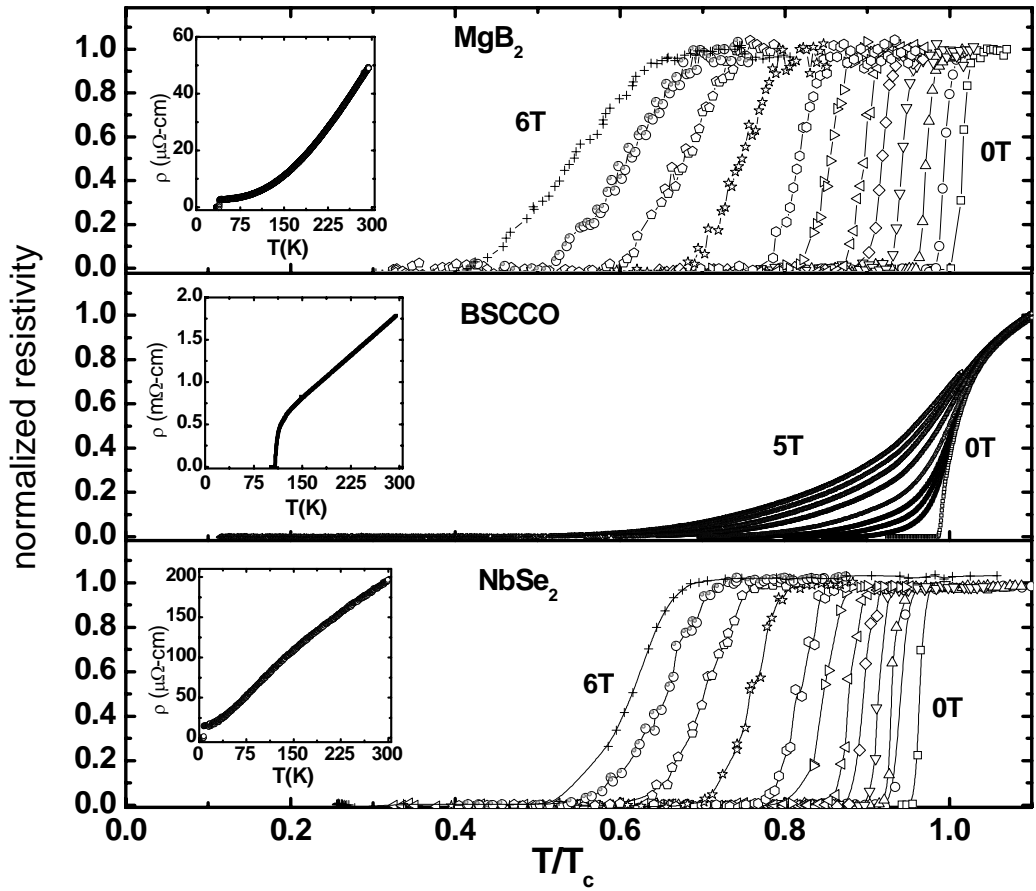

**Figure 2**

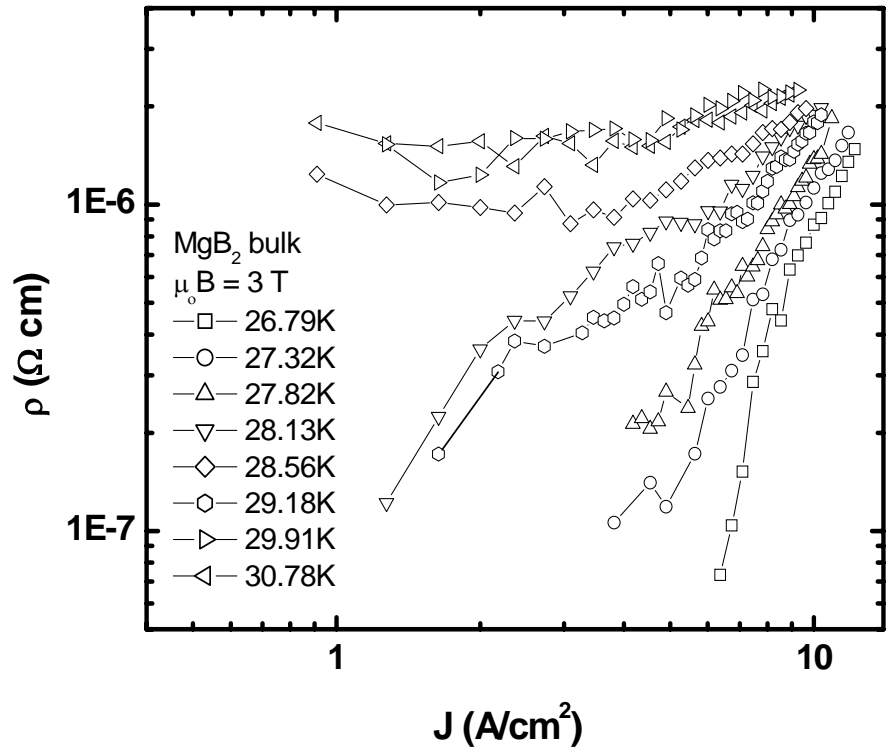

**Figure 3**

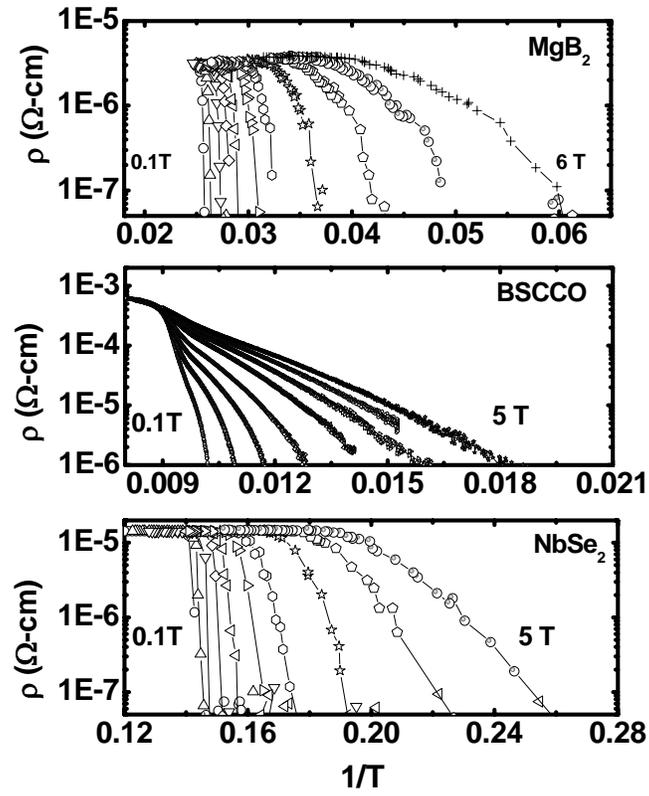

**Figure 4**

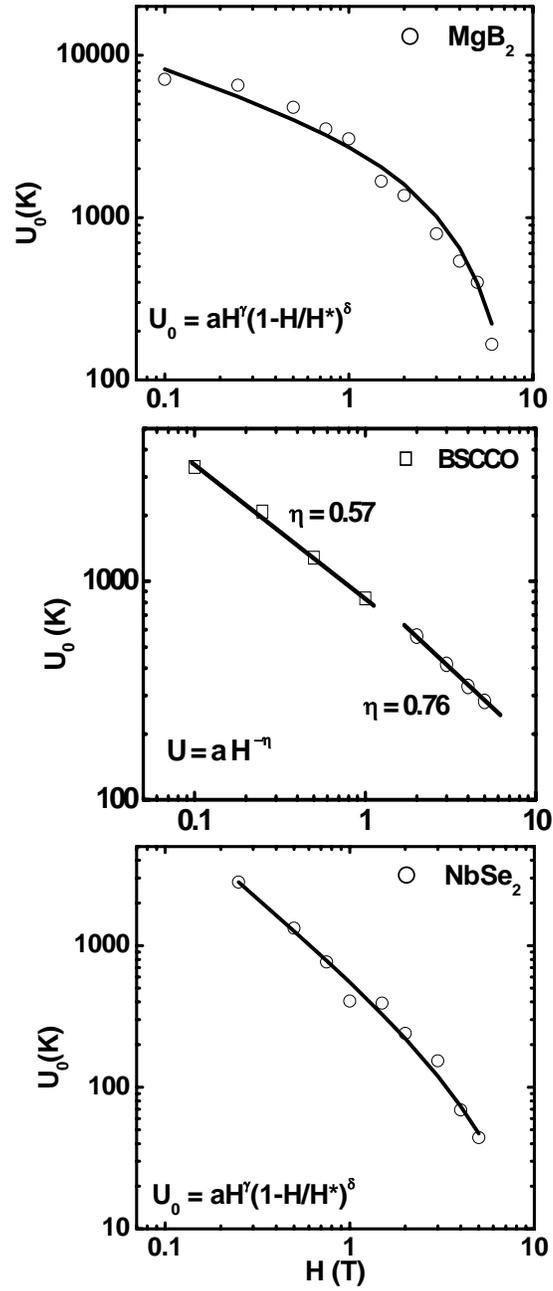

**Figure 5**

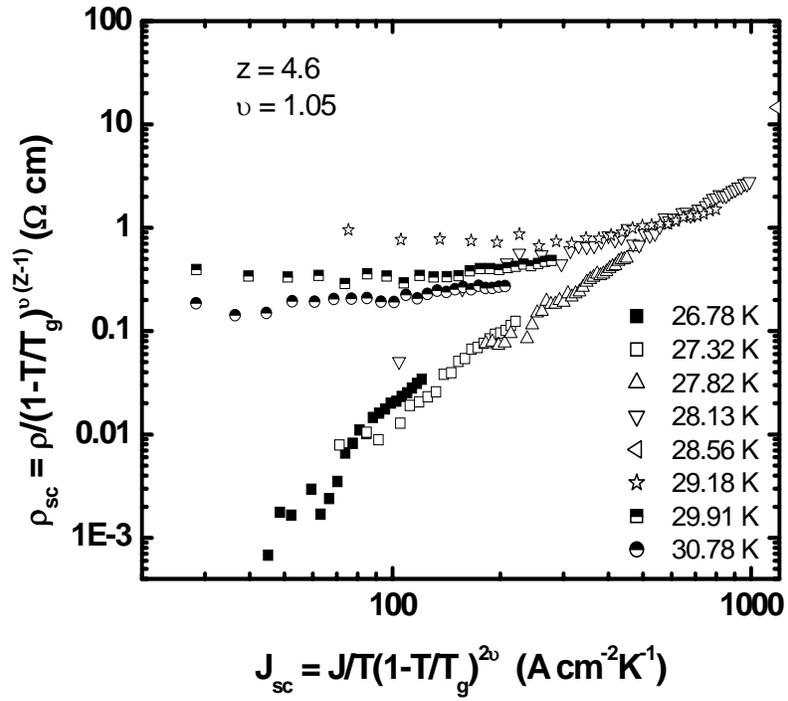